\documentclass[conference]{IEEEtran}
\IEEEoverridecommandlockouts
\usepackage{cite}
\usepackage{amsmath,amssymb,amsfonts}
\usepackage{graphicx}
\usepackage{textcomp}
\usepackage{xcolor}
\usepackage{hyperref}
\def\BibTeX{{\rm B\kern-.05em{\sc i\kern-.025em b}\kern-.08em
    T\kern-.1667em\lower.7ex\hbox{E}\kern-.125emX}}
\begin{document}

\title{Unmanned Aerial Vehicles in Smart Agriculture: Applications, Requirements and Challenges\\
}

\author{\IEEEauthorblockN{Praveen Kumar Reddy Maddikunta$^1$, Saqib Hakak$^2$, Mamoun Alazab$^3$, Sweta Bhattacharya$^1$, \\Thippa Reddy Gadekallu$^1$, Wazir Zada Khan$^4$, and Quoc-Viet Pham$^{5*}$}
\IEEEauthorblockA{$^1$\textit{School of Information Technology, Vellore Institute of Technology, Vellore, India},\\ (Email: \{thippareddy.g, praveenkumarreddy, sweta.b\}@vit.ac.in)}	
\IEEEauthorblockA{$^2$\textit{Faculty of Computer Science, University of New Brunswick, Fredericton, Canada (e-mail: saqibhakak@gmail.com)}}
\IEEEauthorblockA{$^3$\textit{College of Engineering, IT and Environment, Charles Darwin University, Casuarina, Australia (e-mail: alazab.m@ieee.org)}}
\IEEEauthorblockA{$^4$\textit{Faculty of CS \& IT, Jazan University, Saudi Arabia (e-mail: wazirzadakhan@jazanu.edu.sa)}}
\IEEEauthorblockA{$^5$\textit{Research Institute of Computer, Information \& Communication Pusan National University, Busan, Korea} \\ (e-mail:vietpq@pusan.ac.kr$^*$)}
}
\maketitle

\begin{abstract}
In the next few years, smart farming will reach each and every nook of the world. The prospects of using unmanned aerial vehicles (UAV) for smart farming are immense. However, the cost and the ease in controlling UAVs for smart farming might play an important role for motivating farmers to use UAVs in farming. Mostly, UAVs are controlled by remote controllers using radio waves. There are several technologies such as  WiFi or ZigBee that are also used for controlling UAVs. However, Smart Bluetooth (also referred to as Bluetooth Low Energy) is a wireless technology used to transfer data over short distances. Bluetooth smart is cheaper than other technologies and has the advantage of being available on every smart phone. Farmers can use any smart phone to operate their respective UAVs along with Bluetooth Smart enabled agricultural sensors in the future. However, certain requirements and challenges need to be addressed before UAVs can be operated for smart agriculture-related applications. Hence, in this article, an attempt has been made to explore the types of sensors suitable for smart farming, potential requirements and challenges for operating UAVs in smart agriculture. We have also identified the future applications of using UAVs in smart farming.
\end{abstract}

\begin{IEEEkeywords}
Unmanned Aerial Vehicles (UAVs), Smart Agriculture, Bluetooth Low Energy, Smart Farming, Smart Sensors, Agriculture Sensors, Precision Agriculture, Smart Fields and Crops Monitoring.
\end{IEEEkeywords}

\section{Introduction}
\label{Sec:Introduction}

To meet the huge demand for food for the growing population, agriculture has to be revolutionized by using Information and Communication Technologies (ICT). Smart agriculture is the buzzword nowadays which leverages the latest techniques of ICT to grow the food in a sustainable and clean manner \cite{ayaz2019internet}. ICT technologies like the Internet of Things (IoT), Remote Sensing and Unmanned Aerial Vehicles (UAV) \cite{garg2018uav,uddin2020amateur} can effectively use sensors for smart agriculture. Using these technologies, the farmers can automate the irrigation process by remotely monitoring the crop field with the help of several sensors \cite{bu2019smart}. Smart agriculture not only helps in catering the food needs of the growing population, but also helps in recent trends in agriculture like organic farming (Food Shortage, Covid-19 self-sufficient farming). UAVs or drones are a kind of aircrafts that can fly autonomously which can be controlled remotely. ICT tools and techniques like embedded systems, Global Positioning Systems (GPS) and sensors are used to control UAVs. UAVs can be used to monitor situations/missions which are very dangerous and risky for humans. UAVs were originally invented for usage in military applications. Gradually UAVs were successfully employed in several civil applications like agriculture, policing, surveillance, recreational purposes, etc \cite{mozaffari2019tutorial}. Recently e-commerce giants like Amazon started to deliver their products to their customers through UAVs \cite{moshref2020truck}. UAVs have  also been successfully used for monitoring the civilian movements and social gatherings, make announcements in rural or remote areas about the rules of lockdown to combat the Covid-19 pandemic, thereby reducing the risk of police and health authorities being infected by the disease \cite{chamola2020comprehensive}.

\begin{figure*}[tb!]
	\centering
	\includegraphics[width=0.90\linewidth]{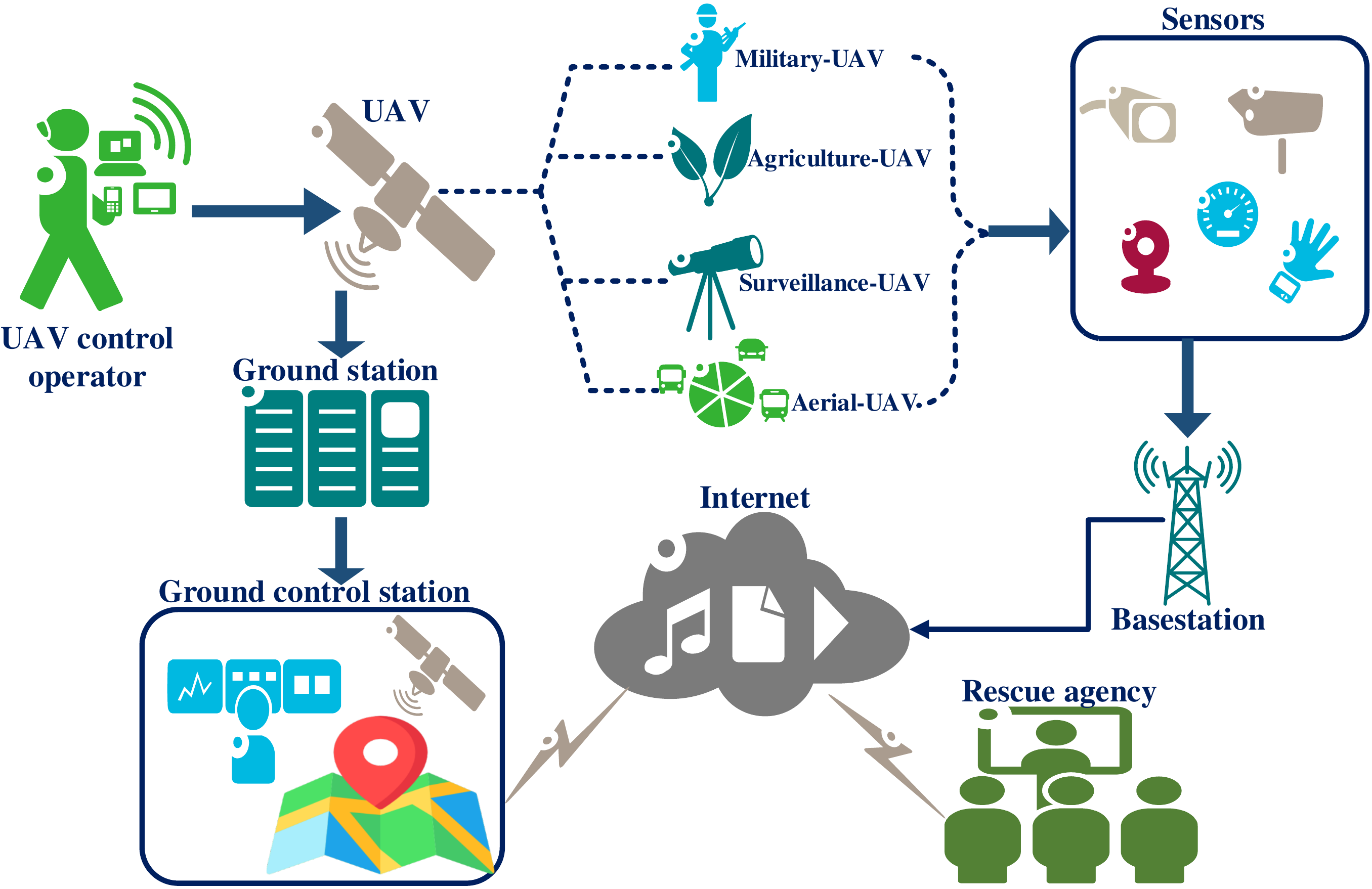}
	\caption{UAV monitoring system.}
	\label{fig:Main Arc}
\end{figure*}

The main advantages of using UAVs for smart agricultural applications include mobility of UAVs in variable weather conditions, ability to capture high-resolution pictures from different ranges (average range 50 to 100 meters) \cite{Pham2020ASurvey_MEC}. It is also possible to use UAVs for determining and monitoring the quality of crops, monitoring attacks attempted by pests/weeds/animals. The farmers and other stakeholders can access the data gathered through UAVs from cloud-based platforms remotely through apps from their smart devices which can help in predicting the yield of the crop, requirements like pesticides, fertilizers,  seed sowing, etc.
There are a few survey articles on UAV and agriculture. In one of the recent studies \cite{boursianis2020internet}, the authors have explored the role of IoT and agricultural UAV in smart agriculture. The main emphases have been laid upon fundamental aspects of IoT technologies including intelligent sensors, IoT sensor types, networks and protocols, and how IoT can be integrated with UAV for the purpose of smart farming. Similarly, in \cite{kim2019unmanned}, the authors have explored the potential applications of using UAV in agriculture and discussed different categories of UAV platforms. Compared to above-mentioned studies, the  main contributions of this article are summarized as follows:
\begin{itemize}
	\item Overview of Bluetooth Smart and types of UAVs and agricultural sensors in smart agriculture are explored.
	\item Types of Agricultural sensors are enumerated with applications and features highlighted
	\item  Enabling Requirements for successful adaptation and usage of UAVs in smart agriculture are enlisted.
	\item Case studies involving  Bluetooth Smart enabled agricultural sensors and UAVs in smart agriculture are explored.
	\item Potential Future Challenges along-with research directions are also presented.
	
\end{itemize}

The rest of the paper is organized as follows. Section~\ref{Sec:Fundamentals} presents a detailed overview on the architecture of UAV, different types of agricultural sensors used in UAVs, state-of-the-art on recent applications of UAVs in several domains.  Section~\ref{Sec:UseCases} discusses potential applications of UAVs in smart agriculture and case studies on real-time implementation of UAVs for smart agriculture. A detailed analysis of the requirements for UAVs for smart agriculture is provided in Section~\ref{Sec:Review}. Challenges and Future research directions are discussed in Section~\ref{Sec:Challenges} and Section~\ref{Sec:Conclusion} concludes the paper.
\section{Background}
\label{Sec:Fundamentals}
UAVs are one type of aircraft that can fly autonomously in the air without the involvement of a pilot on board, and the aircraft's motion is controlled remotely by an operator. UAV consists of sensors \cite{numan2020systematic,patel2020review}, cameras that record and relay images to the operator which is depicted in Fig.~\ref{fig:Main Arc}. UAVs were initially widely used in military applications and surveillance applications. Later, due to the rapid growth of technology in smart agriculture, UAVs have been commonly used in agriculture, helping farmers with crop monitoring, crop spraying, weed detection, disease detection, etc. Some of the advanced features, such as low maintenance costs, fast set-up time, low acquisition costs, and live data capture, have made UAVs a better option for farmers in the agricultural sector \cite{boursianis2020internet}. This section deals with the general architecture of UAVs in agriculture, agricultural applications using UAVs and UAV sensors.

\begin{table*}[t]
	\centering
	\caption{Overview of Bluetooth Technology \cite{B1}.}
	\label{tab:table1}
	\begin{tabular}{|c|c|c|c|c|c|c|}
		\hline
		Version &
		Range &
		\begin{tabular}[c]{@{}c@{}}Power   \\ Consumption\end{tabular} &
		Radio Frequency &
		\begin{tabular}[c]{@{}c@{}}Connection \\ Time\end{tabular} &
		\begin{tabular}[c]{@{}c@{}}Distance\\  Range\end{tabular} &
		Data Speed \\ \hline
		Version 1.X & 10 meters  & 100mW       & 2400-2483.5MHz & $>$100ms   & 100 meters & 1Mbps  \\ \hline
		Version 2.X & 30 meters  & 10mW        & 2400-2483.5MHz & 50-100ms & 101 meters & 3Mbps  \\ \hline
		Version 3.0 & 30 meters  & 2.5mW       & 2400-2483.5MHz & 50ms     & 102 meters & 24Mbps \\ \hline
		Version 4.X & 100 meters & 1mW         & 2400-2483.5MHz & 6ms      & 50 meters  & 24Mbps \\ \hline
		Version 5.X & 240 meters & 0.01–0.50mW & 2400-2483.5MHz & 3ms      & 200 meters & 48Mbps \\ \hline
	\end{tabular}
\end{table*}

\subsection{Overview of Bluetooth Smart}
Bluetooth Smart or Bluetooth Low Energy (BLE) is a wireless system used for technical advancement in the medical, environmental, safety, and energy sectors \cite{antonioli2020key}. Compared to classic Bluetooth technology, BLE requires low power and reduced costs, even if the communication range remains the same. BLE merged into the main Bluetooth standard in July 2010, when Bluetooth Core Specification 4.0 was adopted as Bluetooth Smart and included the classic Bluetooth protocol, Bluetooth High-Speed Protocol and BLE \cite{fraga2020design}. Bluetooth Smart is a dual-mode device, typically a laptop or smartphone whose hardware is compatible with both classic and BLE devices. Bluetooth Smart is a low-energy device that normally has a battery-operated sensor that requires a smart device to operate. BLE is a Bluetooth 4.0 subset with a new protocol stack, designed for very low-power applications that can run off a coin cell battery for months or even years. 
The main features of different versions of BLE are presented in  Table~\ref{tab:table1}.

\subsection{Types of Unmanned Aerial Vehicles}
Here different types of UAVs \cite{ezuma2019detection} are discussed.
\subsubsection*{Multi Rotor UAVs}
These UAVs are mostly used for applications like aerial surveillance, photography etc. These are easy to manufacture and are the cheapest of all kinds of UAVs. 
Different categories of Multi Rotor UAVs are Tricopters which have 3 rotors, Quadcopter which have 4 rotors, hexacopter which have 6 copters and octocopter with 8 rotors \cite{xia2020adaptive}. Some of the limitations are limited flying time of 30 minutes, limited speed, so they are not suitable for projects such as long distance surveillance and aerial mapping.  

\subsubsection*{Fixed wing UAVs} These UAVs are controlled autonomously without a human pilot on-board.  They have average flying time of 2 hours and some of the recent Fixed Wing UAVs can fly upto 16 hours. They are ideal for long distance operations. Some of the limitations are high costs and highly skilled training to operate. They need runway for launching \cite{panagiotou2020aerodynamic}.

\subsubsection*{Single Rotor UAVs}
These category of UAVs look very similar to the helicopters. These UAVs have only one huge rotor and a smaller one near the tail of the UAV. They can fly for higher amount of times compared to multi-rotor UAVs. 
Some of the limitations are more complex and prone to operational risks, higher costs.

\subsubsection*{Hybrid Vertical Take-off and Landing (VTOL)} 
These UAVs are a hybrid of fixed wing UAVs and rotor based models. These UAVs are equipped with sensors and can be controlled remotely \cite{ryi2020study}.
Table~\ref{tab:Table2} exhibits commonly used agricultural UAVs.

\begin{table*}[]
	\centering
	\caption{Types of Smart Agriculture UAV's with Specifications.}
	\label{tab:Table2}
	\resizebox{\textwidth}{!}{%
		\begin{tabular}{|c|c|c|c|c|c|}
			\hline
			\textbf{Agriculture UAVs} &
			\textbf{UAV Type} &
			\textbf{Potential Application} &
			\textbf{\begin{tabular}[c]{@{}c@{}}Nominal \\ Coverage\end{tabular}} &
			\textbf{Type of Sensor} &
			\textbf{Specifications} \\ \hline
			eBee SQ &
			Fixed wing &
			\begin{tabular}[c]{@{}c@{}}RGB imagery, spanning vast \\ areas of every flight, \\ Soil Temperature\end{tabular} &
			500 acres &
			Parrot Sequoia &
			1.1 kg, 43.3 inches \\ \hline
			Sentera PHX &
			Fixed wing &
			\begin{tabular}[c]{@{}c@{}}weed management, pest \\ management, crop health \\ monitoring\end{tabular} &
			700 acres &
			NDVI, NIR, GPS &
			1.8 kg, 45 mph \\ \hline
			Lancaster 5 &
			Fixed wing &
			\begin{tabular}[c]{@{}c@{}}Plants counting and number, \\ assessing plant quality, \\ creating prescription maps\end{tabular} &
			300 acres &
			\begin{tabular}[c]{@{}c@{}}parrot sequoia+, hyper\\ spectral, multispectral, \\ thermal/IR\end{tabular} &
			3.55 kg \\ \hline
			HoneyComb &
			Fixed wing &
			\begin{tabular}[c]{@{}c@{}}Navigating, surveillance, \\ Soil H20 levels, air pressure\end{tabular} &
			600 acres &
			RGB,NIR camera &
			\begin{tabular}[c]{@{}c@{}}3x 8000mAh \\ batteries\end{tabular} \\ \hline
			AgEagle RX-60 &
			Fixed wing &
			\begin{tabular}[c]{@{}c@{}}Aerial Imaging, crop health \\ monitoring, maps prescription\end{tabular} &
			400 acres &
			NDVI Sensor,GPS &
			\begin{tabular}[c]{@{}c@{}}42 mph, 3.17 kg,  \\ 5500 mAh\end{tabular} \\ \hline
			Dji matrice 600 Pro &
			Multi rotor &
			\begin{tabular}[c]{@{}c@{}}Plants counting, Navigating, \\ aerial photography\end{tabular} &
			50 to 100 acres &
			\begin{tabular}[c]{@{}c@{}}parrot sequoia+, GPS, \\ micasense rededge-mx\end{tabular} &
			\begin{tabular}[c]{@{}c@{}}6 kg, 17.9 mph, \\ 5 km range\end{tabular} \\ \hline
			Dji matrice 210 &
			Multi rotor &
			\begin{tabular}[c]{@{}c@{}}Firefighting, pipeline \\ inspection\end{tabular} &
			50 to 100 acres &
			\begin{tabular}[c]{@{}c@{}}FPV Camera, LiDAR \\ scanner\end{tabular} &
			3.80 kg, 7 km range \\ \hline
			Sentera NDVI &
			Multi rotor &
			\begin{tabular}[c]{@{}c@{}}crop health monitoring, \\ Plants counting\end{tabular} &
			50 acres &
			LIDAR sensors, GPS &
			\begin{tabular}[c]{@{}c@{}}1.9 kg, 45 mph, \\ 1,000 ft/min\end{tabular} \\ \hline
			AgBot &
			Multi rotor &
			\begin{tabular}[c]{@{}c@{}}plant height, assessing \\ plant quality\end{tabular} &
			75 acres &
			\begin{tabular}[c]{@{}c@{}}multispectral, \\ thermal/IR\end{tabular} &
			\begin{tabular}[c]{@{}c@{}}4.7 kg, 6500 mAh, \\ 38.4 mph\end{tabular} \\ \hline
		\end{tabular}%
	}
\end{table*}

\subsection{Types of Agricultural Sensors}
In this section, sensors suitable for smart farming are explored and discussed \cite{malik2020leveraging}:

\subsubsection{Location-based Sensors}

Location sensors are used for locating different areas and spots in the agriculture fields \cite{lee2002activity,bayrakdar2020employing}. The farmers utilize various locations sensors that help them during the different stages of the life cycle of crops. Normally, GPS receivers are used for finding the longitude and latitude of a particular point on the earth's surface with the help of a GPS satellite network. These smart location sensors play an important role in precision agriculture by pointing out the location in fields of monitoring growing crops for watering, fertilization, and treatment of weeds.    

\subsubsection{Electrochemical Sensors} These sensors are used to extract a composition from a particular biological sample such as plants, soil etc. \cite{salam2020internet}. In smart agriculture, these sensors are generally used to detect pH levels and soil nutrient levels where sensor electrodes detect specific ions within a soil. 
\subsubsection{Temperature and Humidity Sensors}

Temperature and humidity are among the most important weather factors which directly affect the health and growth of all types of crops. Correct measurement of these environmental factors is helping the farmer adjust the quantity of fertilizer and water \cite{singh2020leveraging}. Various types of temperature, humidity sensors are available which helps the farmers to measure and monitor the levels of humidity and temperatures of their fields and greenhouses. These sensors are wireless-enabled and battery operated.  

\subsubsection{Optical Sensors} These sensors work on the principle of converting light rays into an electrical signal \cite{alvar2020testing}. There are several optical sensors (such as RGB camera, converted near-infrared camera, six-band multispectral camera, high spectral resolution spectrometer etc) that have been used in UAVs for precision agriculture related applications \cite{von2015deploying}. A brief description of few sensors working on this principle are as described as the following.

\textbf{ (a) Visible Light Sensors (RGB):}
Visible Light Sensors (RGB) are most popularly used by UAVs in precision agriculture and related smart agro applications.  It is a recognized fact that human eye is sensitive to red, green and blue bands of light. The RGB sensor in UAV camera captures the image such that they reproduce the same effect as seen with a human eye. Also, the costs of the RGB sensor based cameras are relatively affordable, light weight and extremely good at creating orthomosaic maps that captures images and aerial videos of the entire field at a single instance. This enables to take quicker observations and after entering the geographical data into the GPS, one can immediately get into the root of the problem without affecting the entire field \cite{singh2020odysseys}. The RGB cameras thus help in detailed inspection of agricultural assets efficiently and effectively in varying weather conditions. The associated challenges of this sensor include its inadequacy to analyse large number of agro parameters requiring spectral information existing in non-visible spectrum. 

\textbf{(b) Multi-spectral Sensors:}
Multi-spectral sensors are extremely appropriate for UAV based agricultural analytics. These sensors capture images with exceptional spatial resolution and also possess the capability to determine reflectance in near infrared \cite{nhamo2020assessment}. Thus these sensors are very effective and essential for farmers, researchers and agronomists. The collection of multi-spectral data is an absolute necessity for performing analysis of crop health. The multiple bands of light enable researchers to conduct precision analytic and produce insights on plant vigor, canopy cover, leaf and various other parts of the plants. The absence of such multi-spectral data would make early detection of plant diseases, weeds, pests and calculation of vegetative biomass almost impossible. 

\textbf{(c) Hyper-spectral Sensors:}
Hyper-spectral sensors are extremely capable of capturing detailed images in the spectral and spatial range. These sensors are equipped with area detectors that quantify the captured light resulting from the conversion of incident photons into electrons. The conversion is achieved using two sensors – charge coupled device (CCD) sensors and complimentary metal-oxide-semiconductor (CMOS) sensors. The successful use of hyper-spectral sensors in UAV is possible through the availability of pre-built systems constituting of the sensor manufacturer, the UAV manufacturer and the party responsible for system integration at the pre and post flight level \cite{weiss2020remote}. The combinations of all these three aspects ensures commercial success of the hyper-spectral sensors in measuring hundred bands, performing data processing and achieve decision making in agriculture and forestry. 

\subsubsection{Thermal infrared sensors}
Thermal infrared sensors help to capture the temperature of the objects, generates the images and displays the same based on the information collected. Infrared sensors and optical lenses are used in thermal cameras to capture thermal energy. Normally all objects with temperature greater than absolute zero discharge infrared radiation at particular wavelengths in proportionate to their specific temperatures. The thermal cameras detect the radiations relevant to their wavelengths and converts it to grayscale image generating heat in this process \cite{allred2020overall}. There exist thermal sensors capable of generating colored images in which warmer images are presented in yellow color and cooler ones in blue color. Thermal sensors are widely used for many agricultural related applications such as monitoring of various conditions of crops and soil. The applications of these thermal sensors mounted on UAVs are irrigation management/scheduling by calculating the soil and crop water stress, detection/prediction of various crops disease \cite{gadekallu2020novel,reddy2020deep} (e.g., pathogen), mapping soil texture, crops maturity monitoring for harvesting, localization's of tiles and crops yield mapping etc.

Table~\ref{tab:table3} illustrates different types of agriculture sensors. 

\begin{table*}[h!]
	\centering
	\caption{Types of Agricultural Sensors.}
	\label{tab:table3}
	\resizebox{\textwidth}{!}{%
		\begin{tabular}{|c|c|c|c|c|c|}
			\hline
			\textbf{Sensor} &
			\textbf{\begin{tabular}[c]{@{}c@{}}Available agriculture \\ sensors\end{tabular}} &
			\textbf{Power consumption} &
			\textbf{Connection time} &
			\textbf{Data rate} &
			\textbf{Potential applications} \\ \hline
			\begin{tabular}[c]{@{}c@{}}Location-based \\ Sensors\end{tabular} &
			GPS Receiver &
			battery operated &
			NA &
			NA &
			Precision Agriculture Management, Localization \\ \hline
			Optical Sensors &
			Cameras &
			battery operated &
			1 sec &
			5 Mbps &
			\begin{tabular}[c]{@{}c@{}}Precision  agriculture, Soil properties measurement \\ (e.g., moisture, clay)\end{tabular} \\ \hline
			Thermal Sensors &
			Thermal Cameras &
			battery operated &
			1 min &
			1 Mbps &
			\begin{tabular}[c]{@{}c@{}}Irrigation management/ scheduling, Detection/prediction \\ of various crops disease, Mapping soil texture, Crops maturity\\ monitoring, Localization's of tiles, Crops yield mapping\end{tabular} \\ \hline
			\begin{tabular}[c]{@{}c@{}}Temperature and \\ humidity sensors\end{tabular} &
			\begin{tabular}[c]{@{}c@{}}Wireless Temperature \\ and Humidity sensors\end{tabular} &
			battery operated &
			NA &
			NA &
			\begin{tabular}[c]{@{}c@{}}Agricultural fields and greenhouses temperature and \\ humidity monitoring and measurement. Adjustment of \\ fertilizer and water quantity.\end{tabular} \\ \hline
		\end{tabular}%
	}
\end{table*}

\section{Applications and Case studies of UAV in Agriculture }
\label{Sec:UseCases}
\subsection{Potential applications of UAV in Smart agriculture}
We presently belong to the era of modernization movement where new technologies are being embraced in every sphere of life with promises of better yields and efficiencies. Adaption of UAV in agriculture is an approach that reduces manual farming labour, allowing detailed observation of the cultivation field being unnoticed below the coverage of clouds. The additional benefits obviously includes accelerated deployment, capturing of high resolution images in minimal cost yet performing all the activities similar to a piloted high altitude craft ensuring high safety. The UAVs are equipped with sophisticated and specialized sensors that make them immensely powerful in capturing images of high spatial and temporal resolution. These images help in achieving better insights on the farming resources and livestock thereby providing more accurate and consistent data for better decision making \cite{kim2019unmanned, perera2019unmanned}. The various UAV models and relevant sensors used in agriculture are presented in Table~\ref{tab:my-table}. UAV has huge scope of application in agriculture and some of the major implementation areas are discussed in the following section.

\begin{table*}[h!]
	\caption{UAV Models and their Applications in Agriculture.}
	\centering
	\label{tab:my-table}
	\begin{tabular}{|c|c|c|c|}
		\hline
		\textbf{\begin{tabular}[c]{@{}c@{}}Application in \\ Agriculture\end{tabular}} &
		\textbf{UAV Model} &
		\textbf{Crop} &
		\textbf{Sensor} \\ \hline
		\begin{tabular}[c]{@{}c@{}}Sky-Farming and \\ Crop Monitoring\end{tabular} &
		\begin{tabular}[c]{@{}c@{}}Fixed-Wing, Hexacopter \\ and Quadcopter\end{tabular} &
		\begin{tabular}[c]{@{}c@{}}Wheat, Soya, Barley, \\ Oat and Coffee\end{tabular} &
		\begin{tabular}[c]{@{}c@{}}Digital, Hyper-spectral \\ and Multi-spectral camera\end{tabular} \\ \hline
		Precision Agriculture &
		Fixed-Wing, Rotary-Wing &
		\begin{tabular}[c]{@{}c@{}}All types of crops but mainlyCorn, \\ Soya, Wheat, Vineyard \\ Grapes, Potatoe, Sugar, Citrus \\ Orchards, Rice, Pomegranate,\end{tabular} &
		\begin{tabular}[c]{@{}c@{}}RGB, Hyper-spectral, \\ Multi-spectral and Thermal \\ camera\end{tabular} \\ \hline
		Irrigation Management &
		Fixed-Wing and Quadcopter &
		\begin{tabular}[c]{@{}c@{}}Grapes, Mandarin, Peach, Orange, \\ Vineryad, Barley,Almond\end{tabular} &
		\begin{tabular}[c]{@{}c@{}}Digital, Micro Hyper-spectral, \\ Multi-spectral\end{tabular} \\ \hline
		Aerial Mustering &
		\begin{tabular}[c]{@{}c@{}}Fixed-Wing, Hexacopter, \\ Quadcopter\end{tabular} &
		Stock Mustering &
		\begin{tabular}[c]{@{}c@{}}Digital, Multi-spectral and \\ Hyper-spectral\end{tabular} \\ \hline
		Artificial Pollination &
		\begin{tabular}[c]{@{}c@{}}Helicopter, Hexacopter \\ and Quadcopter\end{tabular} &
		\begin{tabular}[c]{@{}c@{}}Rice, Apple, Almonds, \\ Cherries, Pears, Tulipa\end{tabular} &
		\begin{tabular}[c]{@{}c@{}}Wind speed sensor, High-definition \\ digital camera\end{tabular} \\ \hline
	\end{tabular}
\end{table*}

\subsubsection{UAV as Sky-farmers}
The use of UAVs allows farmers to monitor their fields from the sky level proving them a bird’s eye view of the entire cultivation field. The observations achieved from the sky level provide critical insights on irrigation issues, soil variability and pest infestations. Considering the livestock aspect, UAVs help in counting the animals and also perform detailed study on the food patterns. The information collected from the sky levels helps the farmers to detect problems in priority in order to take the most suitable decisions to manage productivity and gain better profit margins. The crop fields are generally extremely large and hence difficult to monitor posing serious challenges to the farmers. The challenges are further aggravated with volatile weather conditions increasing the risk and labour costs of maintaining the fields. UAVs equipped with RGB or thermal imaging sensor based cameras help to eliminate these challenges. Fixed wing monocopter, Hexacopter and Quadcopters are generally used for crop monitoring embedded with multispectral or hyperspectral cameras. The UAV images are processed to generate mosaic images which are aligned for GIS integration to achieve conclusive decisions at the later stages.  UAVs are also used to spray chemicals on large expansion of cultivation land in shorter time span. This type of crop spraying is efficient as it covers the plant and the soil and protects farmers from being exposed to harmful chemicals. In this case of infrared thermal imaging sensors play a significant role in the evaluation of droplet deposition to ensure uniformity in the field during spraying. Spot spraying is also done in the similar way targeting weeds. The high resolution hyperspectral, multispectral, near infrared or colour infrared sensor based cameras identify the weed positions and jet sprays the herbicide \cite{yinka2019sky, nadal2017urban}. 

\subsubsection{UAVs in Precision Agriculture}
UAV is quite popularly used in precision agriculture to monitor health of the crops using remote sensing technologies and image analytics. In the traditional process of applying remote sensing technique, the images are captured by satellites and manned aircrafts. The images captured are generally very expensive to be used by common farmers yet their resolutions lag quality. On the contrary, small UAVs which are popularly known as drones, act as feasible solution providing superior quality images with the help of the hyperspectral and multispectral cameras. These images are used to derive the vegetation indices enabling farmers to monitor the crop variability and other exceptional conditions. The Normalized Difference Vegetation Index (NDVI) helps to extract information on biomass levels which further helps in achieving useful insights on possibilities of crop diseases, pest infestation, nutrient deficiencies and various other aspects affecting productivity. Fixed wing UAV and rotary wing UAV are both used in precision agriculture embedded with Multipectral or thermal infrared imaging cameras \cite{radoglou2020compilation}.  

\subsubsection{UAV in Irrigation Monitoring}
Irrigation is one of the most important aspects in agriculture and hence efficient irrigation management is extremely necessary for better productivity. One of the major issues in irrigation management is the lack of adequate and accurate data for the deployment of best practices in irrigation management. The use of UAVs helps in acquiring the important irrigation data at any point of time investing minimum costs. Instead of the conventional UAVs, micro – UAVs seem to be more appropriate in collecting high resolution image while flying in lower altitude. In irrigation monitoring, digital camera, RGB cameras, infrared cameras could be potentially used wherein the focus of the camera is set to infinite but the aperture sensitivity and shutter speed are calculated by conducting flight acquisition tests to achieve the best sharpness and light saturation of the image \cite{waskitho2015unmanned, long2016row}. 

\subsubsection{UAV in Aerial Mustering}
The process of mustering involves use of aerial vehicle to locate and gather farming animals feeding across large span of land. It is basically an automation of the traditional job performed by sheep dogs or cowboys. The use of helicopters are quite common in case of extensively larger span of land, but maneuvering the craft yet maintaining the desired level of agility involves extensive training, certification, fuel costs and exposure to risk leading to casualty. UAVs have immensely helped to solve the purpose and have been succesfully used for mustering in Australia and New Zealand. These UAVs are equipped with sirens that herd the farming animals and also guide them towards milking, feeding and shelter areas. These are also commonly known as Air Shepherd drones which uses fixed wing UAV embedded with infrared cameras \cite{yinka2019sky}. 

\subsubsection{UAV is Artificial Pollination}
The decrease in honeybee population has gained immense concern worldwide. Hence the advent of robotic pollinator has gained momentum. The National Institute of Advanced Industrial Science and Technology (AIST) have developed a mirco UAV that would help in artificial pollination using robots. These robots use an absolute innovative methodology of carrying pollen - animal hair coated with gel to transport the pollens. These UAV robots are equipped with with cameras, GPS and involves use of AI technologies. The wind power generated from the UAVs are also used to conduct artificial pollination. However, the force of the wind generated by UAV have been observed to asymmetrically disperse the pollen which acts as a challenge and future direction of research \cite{chechetka2017materially,li2015distribution, jiyu2017distribution}. 

\subsection{Case studies}

\subsubsection{Use Case 1: Renewable Energy based UAV for Agriculture}
\begin{figure}[t]
	\centering
	\includegraphics[width=\linewidth]{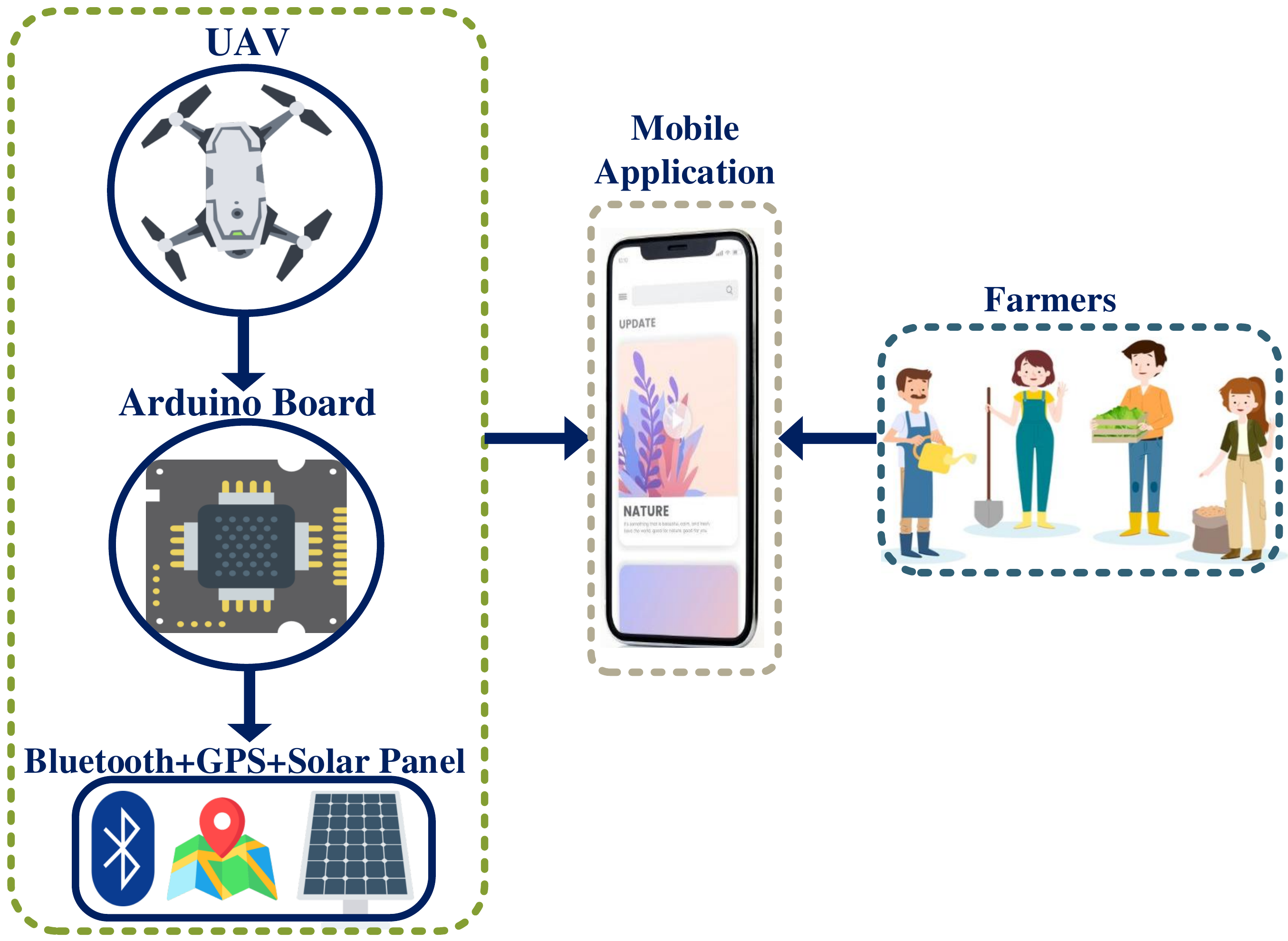}
	\caption{Renewable energy based UAV in smart farming.}
	\label{fig:case1}
\end{figure}

The first project is a potential Bluetooth embedded UAV having significant potential of application in agriculture. The case study emphasizes on importance of uniformly disseminating pesticides and fertilizers in cultivation fields using UAV. As presented in Fig.~\ref{fig:case1}, the READ pesticide spraying hexacopter is used for spraying in the cultivation fields in this case reducing significant manual labour and work load ensuring optimum security for the farmers. The farmer basically can control the mirco - UAV or drone using an android application with the help of a Bluetooth module. The Bluetooth module helps the farmer get connected to the app that is interfaced in the drone. Drones or UAVs normally operate remotely wherein the operator focuses on the visual contact with the aircraft or maneuvers the craft through pre-programmed paths using GPS. The craft typically follows the route of the cultivation land using GPS. The Arduino board being an prototype open source electronics platform is embedded or interfaced with the Bluetooth module and the GPS. The other important aspects of being aerially stable, balanced and oriented are managed by the use of accelerometer, magnetometer and gyro. To achieve energy efficiency instead of using bio-fuel and hydrogen fuel cells, solar technology is a potential solution which would increase durability yet ensure the craft is light weight to maintain its agility. Hence installation of solar panels would provide additional power and increase flight time for the UAV. If the durability, flight time and power is enhanced the coverage of land area by the UAV would automatically get increased. Thus the above mentioned UAV framework would definitely act as a potential UAV solution with higher efficiency, reliability in reduced cost. This Bluetooth embedded UAV has potential applications in crop monitoring, precision agriculture, spraying, irrigation management and crop planting 
\cite{ezuma2019detection}.

\subsubsection{Use Case 2: Bluetooth Embedded UAV for Air traffic Control}
\begin{figure}[t]
	\centering
	\includegraphics[width=\linewidth]{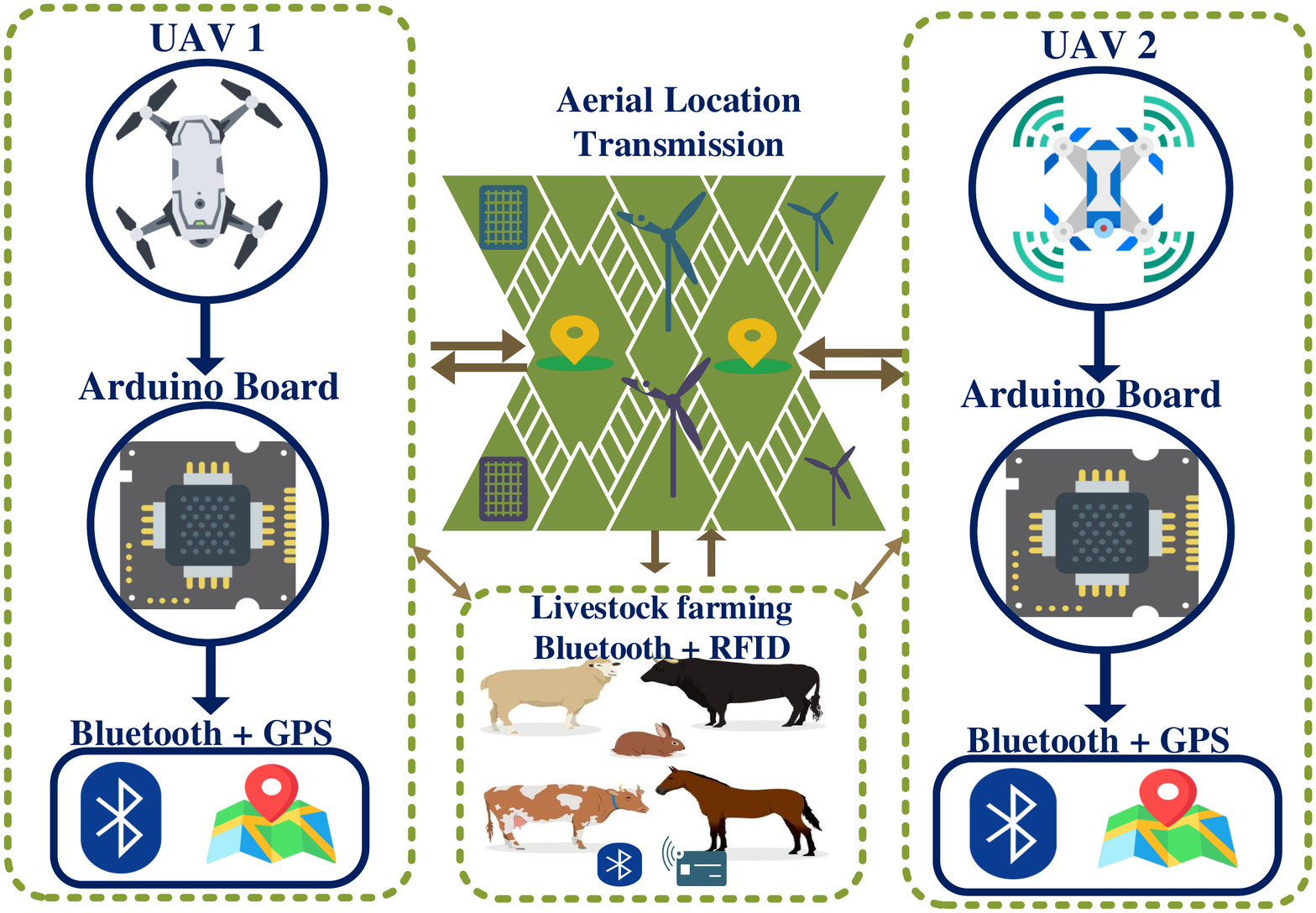}
	\caption{Bluetooth enabled UAV for Livestock Monitoring.}
	\label{fig:case2}
\end{figure}

With the increasing popularity of UAV adaption in agriculture and other domains, there will eventually arise a situation where UAVs will start competing for space and the emancipating experience of flying UAVs will perish. The result of such aerial congestion would be disastrous leading to unprecedented collisions and quadcopter damages with huge loss of infrastructural resources. To eliminate such issues, Intel has come up with a cutting edge Bluetooth technology which would enable UAVs to broadcast their specific aerial location so that other similar crafts or devices can maintain safe distancing in real time. As depicted in Fig.~\ref{fig:case2}, the safety protocol installed in this framework ensures constant communication between the UAVs thereby transmitting static and dynamic frequencies at different frequency levels. The quadcopters are tagged with unique IDs to allow seamless tracking and the location data is transmitted to the connected application. The adaption of Bluetooth technology would ensure accurate detection of the aerial device to the range of almost 2,625 feet in low cost with efficient implementation \cite{U2}. However, a major challenge of such implementation in crowded locations would be difficult considering the receding strength of Bluetooth signal due to obstructions. The falcom like structured UAVs embedded with Bluetooth developed by Intel has the aforementioned Bluetooth empowered safety feature which enables wireless communication between quadcopters and also shares data pertinent to speed, altitude, direction, model between UAVs. This type of UAVs are predominantly used in live stock management. With the advancement in farming, Bluetooth technology based identification tag could be attached to the cattle and the same could be monitored with the help of Bluetooth equipped quadcopters to find exact location of strayed lost cattle. With the help of similar tags, the habits and movements of livestock could also be monitored using the Bluetooth based safety feature in UAVs. These UAVs when connected with android apps could provide alert at the beginning of calving and also when there is an exceptional amount of delay in the calving process generating an alert to introduce timely interventions. Also in case of attacks by ferocious predators, the same alert system is equally beneficial preventing cattle mortality. 

\subsubsection{Use Case 3: Bluetooth Embedded UAV for Self Sustained Ecosystem}
\begin{figure*}[t!]
	\centering
	\includegraphics[width=1.0\linewidth]{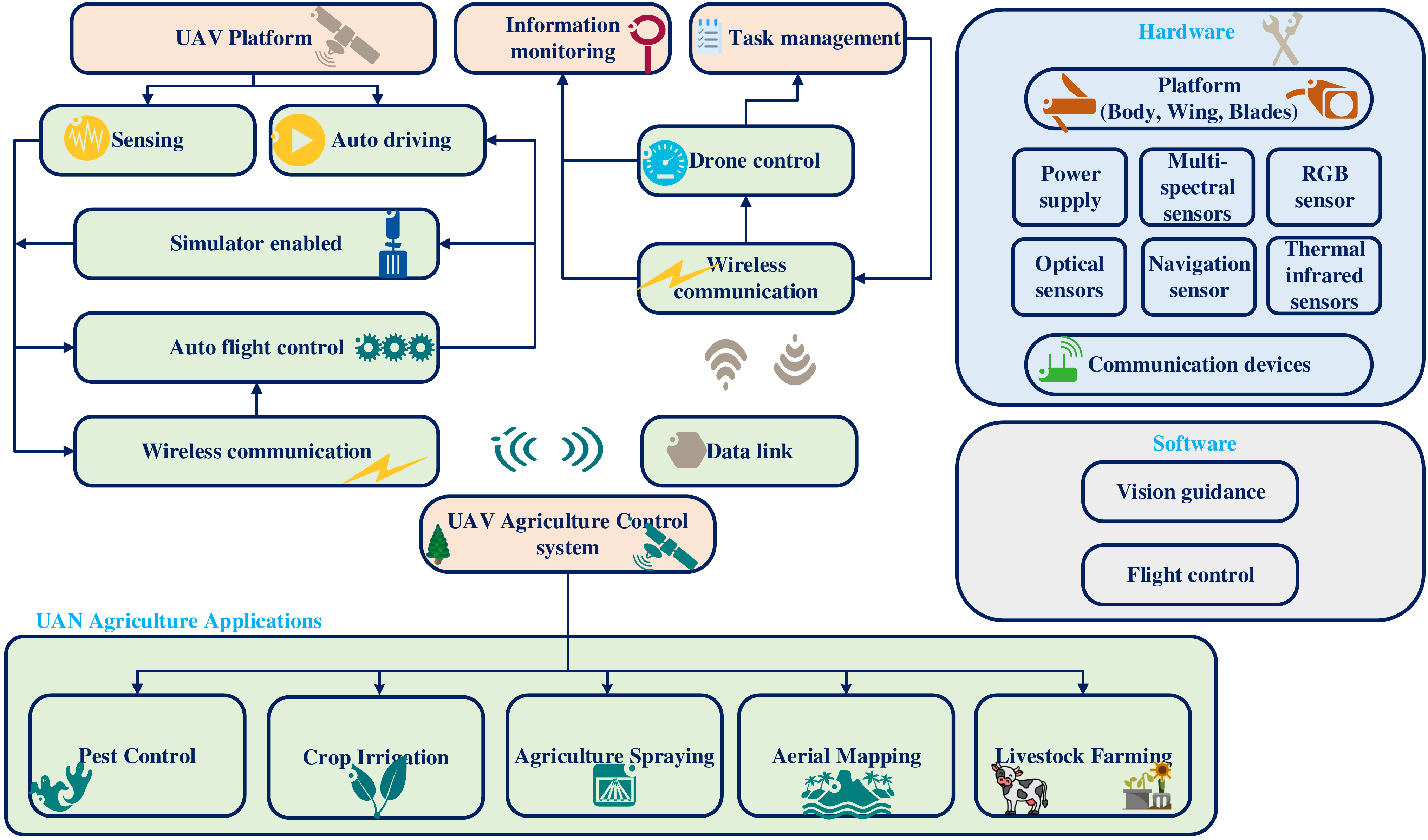}
	\caption{UAV Agriculture Control System.}
	\label{fig:Arc}
\end{figure*}

There has been various successful implementations of UAV in varied sectors of agriculture which supports the basis of human survival and sustenance.  As an example UAV implementations have been predominant in Crop Irrigation, Agriculture Spraying, Aerial Mapping, Livestock Management and also in Pest Control. However the future lies in the integration of all these applications could build a self sustained ecosystem eliminating human interventions. The framework in Fig.~\ref{fig:Arc}. presents a UAV based framework involving the use of an UAV Agriculture control system that would perform all the aforementioned agriculture tasks. The control system would be incorporated with the UAV platform constituting of sensors, auto drivers, simulators, auto flight controls all which would communicate through wireless communication to perform individual tasks. The monitoring of the operational environment and control of the UAV would be conducted using drone controllers and wireless communication with the base stations. To perform all of these activities effectively and efficiently, the hardware to be used to operate the drone and the other devices would play an extremely important role. The framework would include use of various multi-spectral, RGB, optical, thermal infrared users that would help to capture information to instigate actions. The flight control and vision guidance software would be used to monitor and track the vehicular movement and action. The objective would be have a system wherein the drone would automatically perform irrigation as per the soil needs, perform agricultural spraying of seeds and fertilizers, spray pest control specifically on affected regions, conduct aerial mapping and also manage livestock automatically without farmers physical intervention. This would enhance the agro-economic condition of the farmers and contribute towards improvement in agricultural production of the country.

\section{Requirements}
\label{Sec:Review}
In this section, we have highlighted the key requirements of employing UAVs in smart agriculture.

\subsection{Regulation of UAVs}
Regulation of operating UAVs worldwide is one of the fundamental requirements for its successful integration with the smart agriculture. There are many countries where operating UAVs is not allowed due to several reasons. There are few organisations such as, Technical Centre for Agriculture and Rural Cooperation (CTA) which has published certain laws of using UAVs for agricultural purposes but as per \cite{jeanneret2016drone}, still around 73 percent of  African, Caribbean and Pacific countries (ACP) countries does not have any regulation or law with remaining percent having few laws. Similarly, Electronic Communications Committee (ECC) within the European Conference of Postal and Telecommunications Administrations (CEPT) published the report on regulatory aspects of UAVs in Feb 2018 and is under study \cite{fotouhi2019survey}. Hence, until the regulation of using this technology are not published world-wide, it would be quite difficult for the farmer communities to get benefit from the technology. 

\subsection{Network Availability} 
Availability of network is yet other one of the core requirements for using UAV in smart agriculture. Although sending data via Bluetooth Smart from UAV back to the smart phone does not require an internet connection but sending that data to some other platform such as cloud or any storage platform will require strong internet connection and bandwidth. A slight glitch to a strong network or a weak network for real-time application within smart agriculture scenario will have serious consequences for farmer communities. Wireless networking technologies such as 5G and software-defined networking can significantly improve the overall scenario in terms of routing and strong internet connection \cite{Pham2020SumRate,zeng2019accessing}. 

\subsection{Data Storage}
There are numerous applications of utilising UAV in smart agriculture such as precision agriculture, irrigation monitoring etc., as highlighted in section IV. All these applications involves numerous operations such as capturing high-definition images, extracting log-files, analysing data and so on. All these operations require immense amount of data storage especially if the storage is static where there is no internet connectivity and the data needs to be stored within UAV itself. Mostly, the UAVs are equipped with inbuilt storage capacity of 2 gigabytes to 4 gigabytes with a latest quadcopter i.e. DJ1 Mavic Air (comes with 8GB of storage) which might not suffice for storing data related to agricultural activities. Hence, cost-effective and secure data-storage is one of the core requirements for utilisation of UAVs in smart agriculture. Although, cloud-based solutions such as \cite{chen2019uav,luo2015uav,itkin2016development} might be the solution but that also requires strong internet connection which is not always available. Therefore, an alternate low-cost solution based on new technologies such as virtual storage, software-defined storage etc., might be the feasible solution.  
\subsection{Security and Privacy}
The other foremost requirement is securing the data from cyberattacks and maintaining the privacy during the operation. There are number of cyberattacks possible while UAV is performing its required task. Attacks such as WiFi-based including Eavesdropping, denial of service (DoS), information injection etc. can severely affect the whole operation. There are number of studies such as \cite{khan2020uav,horowitz2020cyberattack,manesh2019cyber} where all types of such attacks are identified. With respect to privacy, there are concerns as well such as taking photographs secretly of properties surrounding the agriculture land, spy-related concerns etc. Therefore, tedious efforts are needed to address these concerns. Blockchain-based solutions \cite{hakak2020industrial,hakak2020securing} seems potential solution for addressing the security concern. For addressing privacy concerns, well-laid regulations seems viable option but other solutions also needs to be explored.   

\subsection{Efficient and Low Energy Consumption}

A typical UAV can fly on a stretch for about 15-25 minutes on a single battery \cite{galkin2019uavs}. There are several heavy applications in smart agriculture such as  livestock monitoring, sending data in a real-time for analysis, monitoring soil moisture, weed detection, humidity monitoring  etc. which requires more complex operations. Such complex operations (e.g., long flight-duration, taking high resolution images using infrared, multispectral and hyperspectral sensors etc) will require more processing power and  consume more battery-life of an UAV. Therefore, UAV with high energy-consuming  feature will decrease the overall efficacy of applications within smart agriculture. 

\subsection{User Acceptance of UAV technology}

Acceptability of UAV based technologies in agriculture is often an issue due to the absence of a standardized workflow affecting its popularity. Also, UAVs have shorter flight time ranging between minutes to an hour which does not suffice the requirement for acreage coverage expected by the farmers. The range of the flights are also limited to certain radius per flight time. Secondly, the agricultural UAVs use the same airspace as any general manually operated aircraft which increases chances of inference. With the initial installation cost of UAV being high, if such exceptional incidents occur, the associated financial loss would be enormous. Hence traditional farmers think twice before accepting such technologies. Last but not the least, the images captured by UAVs require analysis by skilled professionals to achieve valuable information. An average farmer often feels intimidated to use such sophisticated technology and would find it difficult to adopt even after training is provided. 

\subsection{Operational Ethics}
Ethical issues in the use of drones are primarily based on two factors - firstly the activities performed by the UAV and secondly the consequential actions performed by the person using the information collected by the UAV. Thus, the action of the UAV can be evaluated depending on the intention and actions performed by the individual controlling the UAV. The use of UAVs are always subject to evaluation unless it is completely automated with absolutely no human interventions. Agricultural sector has always been very open towards acceptance of autonomous operations as the perceived risk factor is quite low. But considerations of occupational health and safety have often been ignored. There is an dire need for regulatory bodies to monitor and control the results  produced by technology. The use of data collected by the UAVs are mostly unregulated and are prone to be collected unethically, used by unauthorized corrupt individuals or hacks for the fulfilment of selfish personal gain. This could lead to facilitation of illegal activities providing information about wild life, cultivation and law enforcement efforts to immoral individuals. Naturally local population would tend to avoid or become hostile towards the use of such technology being deprived of its benefits \cite{sandbrook2015social,fotouhi2019survey}. 

\subsection{Accuracy of Results}
Although farmers have been extremely keen in adopting UAVs in agriculture the accuracy of UAVs data collected in agriculture is often questionable. The use of drones and UAVs allow farmers to monitor the crops from the air effectively. The robotic tractors connected to the UAVs are used to irrigate and fertilize the identified cultivation lands in minimized time frame. This process involves the use of multi-spectral imaging sensors which measures energy reflected from the crops within the specific sections of electromagnetic spectrum. But often the results may not be as reliable as perceived due to the erroneous technique of collecting images from the drones using the multi-spectral sensors. The inconsistencies of the UAV altitude and the angle of sun have significant impact on the results generated, unless controlled by skilled professionals. These inconsistencies have high risks pertinent to generation of erroneous results in comparative analysis of reflections in tree canopies, mild variations of vegetation and  predictions leading expensive consequences for the farmers.

\section{Challenges and Future Research Directions}
\label{Sec:Challenges}
In this section several challenges faced by the proposed BLE enabled UAVs for agriculture are presented that will guide the future research on usage of UAVs in agriculture. 

\subsubsection{Short remote-range of BLE} One of the most interesting challenges is short range of BLE. As specified earlier, the maximum range of latest BLE version is 100 meters which is quite low for a large farm lands. To solve this challenge, a distributed sensing and actuation network of BLE-enabled UAVs can be formed
\cite{motlagh2016low}
where master UAV will be controlled by a farmer directly and other UAVs will act as slaves located at an appropriate distance from each other. However, achieving synchronisation between master and slave UAVs, minimizing latency, achieving healthy data-transfer rate etc. are few of the challenges that requires tedious research efforts.   

\subsubsection{Achieving Higher-Data rates for Dynamic Storage of Data} There are different versions of BLE available with Version-5 having theoretical data-rate of 24 Mbps which is much better than its predecessors. In smart farming, there are several applications such as live-stock monitoring which requires much higher data-rate for data-transfer. As more advancements are coming in BLE versions, it is expected in future BLE-versions with much higher data-rates will be available. Therefore, before utilising BLE-controlled UAVs for smart agriculture, data-rate needs to be enhanced. With data-rate of merely 24 Mbps, dynamic storage of data into cloud or edge-based platforms seems impractical. A single image from high-definition camera using location-based sensors is more than 24 MB in size. Therefore, to store multiple images in a real-time scenario on a Cloud/Edge-based platforms is not possible unless and until higher data-rates are achieved.   

\subsubsection{Interference} BLE uses the frequency range of 2.4 GHz. This range is also used by WiFi, Zigbee and regular Bluetooth technologies. Hence, there is a huge risk of interference and reduced latency when the number of devices operating on this frequency rises. Due to the interference, the devices might get disconnected abruptly or perform poorly. Although farm lands are remotely isolated in rural areas where the possibility of interference is low, this challenge needs to be addressed keeping the concept of future smart cities into consideration where farm lands will be incurred as well.One of the common way of mitigating interference include removing barriers such as metal, concrete, glass etc.

\subsubsection{UAV Technology Acceptance}

The success of any technology depends upon its user's acceptance. Adaptation and accurate usage of high tech and sophisticated technology like UAV's require skills and knowledge. The usage of UAVs by the farmers with limited or no skills is a challenging task. Moreover, successful acceptance of any technology also depends upon the willingness to use it by its consumers. High skill requirements for flying UAVs by the farmers with no or limited flying skills will also affect the willingness of use.  Another factor that may affect the acceptance by UAVs is to ensure the privacy of others while using these UAVs. Ensuring the privacy of others and to avoid any legal implication faced due to privacy violation may also hinder the acceptance of these UAVs in agriculture. Therefore, it is very challenging to encourage and motivate the farmers to accept the UAVs. So there is a need to design and develop effective user acceptance models which identify and provide solutions for ease of use, willingness to use and ensuring the privacy of others for successful adaptation and usage of UAVs in agriculture and to get full benefits from these high tech, sophisticated technologies.

\section{Conclusion} 
\label{Sec:Conclusion}
In this article, the architecture, adaption and usage of UAVs in smart agriculture have been explored and presented. Potential case studies involving Bluetooth Smart-enabled sensors and UAVs in smart agriculture have been discussed. Bluetooth Smart technology can be replaced with any other technology for implementation purposes. The motivation of using Bluetooth Smart in case-studies is the low-cost and ease of access via smart phones. We have also explored various types of agricultural sensors such as location-based sensors, optical sensors, temperature-based sensors, etc, and identified several applications of UAVs in smart agriculture. Besides, we have identified key enabling requirements of UAVs in smart agriculture including acceptance of technology by farmers, accuracy of results, network availability, data storage, regulation of UAVs, and many others. Finally, key research challenges and future directions have been highlighted and discussed.

\bibliographystyle{ieeetran}
\bibliography{references}
\end{document}